# Complex absorbing potential based Lorentzian fitting scheme and time dependent quantum transport

Hang Xie[1*], Yanho Kwok[1], Feng Jiang[2], Xiao Zheng[3] and GuanHua Chen[1**]

[1] *Department of Chemistry, The University of Hong Kong, Hong Kong*
[2] *Department of Physics, Shanghai University of Electric Power, Shanghai, China*
[3] *Hefei National Laboratory for Physical Sciences at the Microscale, University of Science and Technology of China, Hefei, China*
Email: xiehanggm@gmail.com*; ghc@everest.hku.hk**

Based on the complex absorbing potential (CAP) method, a Lorentzian expansion scheme is developed to express the self-energy. The CAP-based Lorentzian expansion of self-energy is employed to solve efficiently the Liouville-von Neumann equation of one-electron density matrix. The resulting method is applicable for both tight-binding and first-principles models, and is used to simulate the transient currents through graphene nanoribbons and a benzene molecule sandwiched between two carbon-atom-chains.

## I. Introduction

As the rapid development of nanotechnology in fabrications and measurements, the nano-electronics becomes an important field in both semiconductor industry and academic research [1,2]. Nano devices such as silicon nanowires, graphene nanoribbons and carbon nanotubes are the subjects of contemporary research. At such small scales, the quantum mechanical effects prevail over the classical behaviors for electron transport.

In the theoretical treatment of nanoscale transport, the nonequilibrium Green's function (NEGF) method has been widely used [3,4]. The density functional theory (DFT) is often combined with NEGF to calculate the transport of the molecules or nano-structures at first principles level [5-7]. For the time dependent quantum transport, the theories are more complicated and the calculation for large systems still meets much challenge [8-9]. Some of the methods focus on the wavefunction propagation [10] and some others focus on the density matrix evolution with the lead spectrum approximation [9] or some time decomposition scheme [11-12].

Recently we developed a new method to calculate the time dependent quantum transport based on the NEGF theory [13-17]. This method, termed as the time dependent density functional theory –nonequilibrium Green's function (TDDFT-NEGF) scheme, treats the lead spectrum exactly, which is beyond the commonly used wide band limit (WBL) approximation [8-9]. Instead of solving the Green's functions directly, we follow the dynamics of dissipation matrices. Together with the density matrix, their equations of motion constitute a close set of equations which can be solved numerically. This method can be employed to simulate any systems in principle. But for the large systems, an effective Lorentzian fitting scheme for the lead self-energy matrix is very difficult. The large number of Lorentzians leads to huge memories and heavy computation load in the TDDFT-NEGF calculation. In our previous paper [16], we proposed several fitting schemes based on the nonlinear least square (LS) method.

However, these fitting schemes may not generate a unique Lorentzian expansion of the self-energy since there are many local minima in the high-dimensional LS parameteric space. In this paper we develop a new Lorentzian fitting scheme based on the complex absorbing potential (CAP) method.

CAP is an effective way to approximate the infinite environment of a finite system of interests [18-22]. It was initially proposed to reduce the reflection of electronic wavefunction at the boundary of a finite region [18]. CAP is also used in quantum transport and reaction dynamics calculations [19-22]. The CAP method is similar to the perfectly matched layer (PML) method which is widely applied in the computational electromagnetics [23]. All these methods introduce some absorbing properties at the boundary regions to reduce the reflection of the wavefunctions. Another advantage of the CAP method is that the Green's function at all energy points can be calculated directly and efficiently, without iterative calculations of the surface Green's function at individual energy point. In this paper we employ the suitable CAP as a practical scheme to derive a unique Lorentzian expansion for the self-energy in the TDDFT-NEGF calculation.

This paper is organized as follows. Sec. II gives the basic theories of our work, such as the Lorentzian expansion in the TDDFT-NEGF calculation; the introduction of CAP and the eigenvector expansion; and a brief introduction to the TDDFT-NEGF theory. In Sec. III, the calculation results and discussions are presented. We show some examples such as the 1D atom chain, the graphene nanoribbons (GNR) and carbon-atom-chain for the CAP calculations. With the Lorentzian expansion from the CAP method, the dynamic quantum transport calculations for these nano-structures are obtained. Sec. IV is the conclusion. Technique details are given in the Appendix.

## II. Theory
### A. Lorentzian expansion

In TDDFT-NEGF theory, the lesser self-energy at the equilibrium state is expressed as follows [13-14]:

$$\overline{\mathbf{\Sigma}}_\alpha^<(\tau-t) = \frac{i}{2\pi}\int_{-\infty}^{+\infty} f_\alpha(\varepsilon)\mathbf{\Lambda}_\alpha(\varepsilon)\cdot e^{-i\varepsilon(\tau-t)}d\varepsilon \tag{1}$$

where $f_\alpha(\varepsilon)$ is the Fermi-Dirac distribution function: $f_\alpha(\varepsilon) = F(\frac{\varepsilon-\mu_\alpha}{k_B T})$ ($k_B$ is the Boltzmann constant, $T$ is the temperature, $\mu_\alpha$ is the chemical potential of the lead $\alpha$), which is expanded by the Padé spectrum decomposition [24]

$$F(z) = \frac{1}{1+\exp(z)} \approx \frac{1}{2} + \sum_{p=1}^{N_p}(\frac{R_p}{z-z_p^+} + \frac{R_p}{z-z_p^-}); \tag{2}$$

and the linewidth function $\mathbf{\Lambda}_\alpha(\varepsilon)$ is expressed by a Lorentzian expansion

$$\mathbf{\Lambda}_\alpha(\varepsilon) \approx \sum_{d=1}^{N_d} \frac{\mathbf{A}_{\alpha d}}{(\varepsilon-\Omega_d)^2 + W_d^2}. \tag{3}$$

With this expansion, the integral in Eq. (1) can be transformed into a residue summation and the equations of TDDFT-NEGF can be recast into a discretized form, which is numerically solvable as detailed in reference [14]. This expansion is called the Lorentzian-Padé decomposition scheme.

As the linewidth function above is a matrix, we have to find a minimal set of Lorentzian functions to fit each of the matrix elements accurately, which is a non-trivial task. The quality of the fitting will determine the accuracy of TDDFT-NEGF calculations, and the number of Lorentzians is related to the computational load. So the Lorentzian fitting is a very important step in our calculations, in particular for the large systems. In our previous paper we proposed several Lorentzian fitting schemes based on the LS method [16]. Because of the large number of fitting parameters, the fitting solution is not unique and there exist many 'local minimum' solutions in the solution space. In the following parts, we show that from the CAP method a universal Lorentzian expansion can be derived.

### B. Complex absorbing potential method

CAP is an artificial potential to mimic the infinite environment by imposing an absorption potential in finite region on the boundary. The commonly-used CAP is derived from the semiclassical approximation by minimizing the reflection coefficient in a 1D quantum wave system [19]. This potential increases from zero on one side of the CAP region near the device to infinity near another side. Figure 1 shows the profile of the CAP. The CAP region consists of a series of repeated blocks in the positions of two leads. One most used CAP has the following form

$$W(z) = i \cdot \frac{\hbar^2}{2m}(\frac{2\pi}{\Delta z})^2 f(z) \tag{4}$$

$$f(z) = \frac{4}{c^2}[(\frac{\Delta z}{z_2 - 2z_1 + z})^2 + (\frac{\Delta z}{z_2 - z})^2 - 2],$$

where $z_1$ and $z_2$ is the beginning and ending position of the CAP region and $\Delta z = z_2 - z_1$ is the length of the region. $c$ is a constant, which is not sensitive to the final result unless it is too large or too small. In this work we set $c=1.0$. After projecting the CAP into the atomic basis ($\{\phi_n(x, y, z)\}$), the following CAP matrix is obtained

$$W_{\alpha,mn} = \int \phi_m^*(x, y, z) W_\alpha(z) \phi_n(x, y, z) dx dy dz. \tag{5}$$

For an isolate system including a device and two CAP regions (left and right), we can calculate its Green's function ($\mathbf{G}_{CAP}^r$), in comparison with the common NEGF calculation for the device's Green's function $\mathbf{G}_D^r$

$$\mathbf{G}_{CAP}^r(E) = \begin{bmatrix} E\mathbf{I}_L - \mathbf{H}_L - \mathbf{W}_L & -\mathbf{H}_{LD} & 0 \\ -\mathbf{H}_{DL} & E\mathbf{I}_D - \mathbf{H}_D & -\mathbf{H}_{DR} \\ 0 & -\mathbf{H}_{RD} & E\mathbf{I}_R - \mathbf{H}_R - \mathbf{W}_R \end{bmatrix}^{-1} \tag{6a}$$

$$\mathbf{G}_D^r(E) = (E\mathbf{I}_D - \mathbf{H}_D - \sum_{\alpha=L,R} \mathbf{\Sigma}_\alpha^r(E))^{-1}, \tag{6b}$$

where $\mathbf{I}_D$ and $\mathbf{I}_L$ and $\mathbf{I}_R$ are the unit matrices with different dimensions; $\mathbf{H}_{DR}$, $\mathbf{H}_{RD}$, $\mathbf{H}_{DL}$, and $\mathbf{H}_{LD}$ are the coupling matrices between the lead and the device; $\mathbf{H}_D$ and $\mathbf{H}_L$ ($\mathbf{H}_R$) are the Hamiltonians of the device and lead regions; $\mathbf{W}_L$ and $\mathbf{W}_R$ are the CAP part in the left and right lead evaluated from Eq. (5); and $\Sigma_\alpha^r(E)$ is the retarded self-energy evaluated from the iteration method [25]. Since CAP mimics the infinite leads, the calculated physical property of the device region (or the device portion of $\mathbf{G}_{CAP}^r$) is very close to that calculated from the NEGF theory (or $\mathbf{G}_D^r$). However, the lead portions of $\mathbf{G}_{CAP}^r$ have no such correspondence with the lead regions in the open system. Only in the positions very close to the device, $\mathbf{G}^r(E)$ of the two systems have close values. Figure 1 shows such correspondence in two systems. The upper panel shows an open system with device and two sets of leads with infinite units; the lower panel shows the CAP case: the device region and two CAP regions with finite units. The imaginary part of CAP is demonstrated by the blue curve. It is noted that $\mathbf{W}_\alpha$ is energy independent, which is much easier to be evaluated than the iterative calculation of $\Sigma_\alpha^r(E)$.

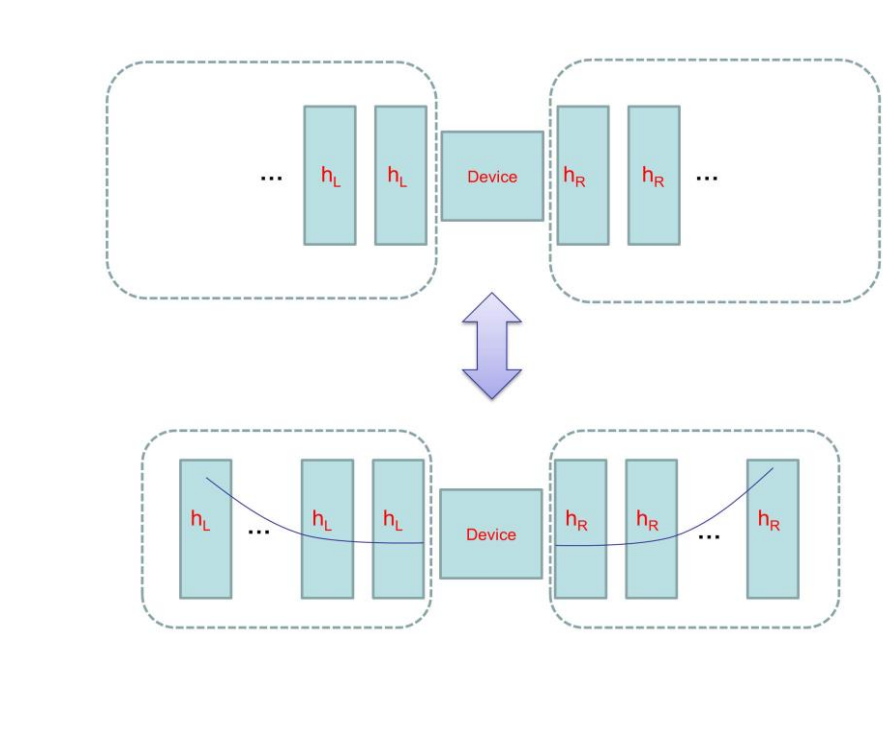

FIG, 1 The demonstration of the CAP method. The upper part shows the common transport case and the lower part shows the CAP scheme for such transport calculation. In the upper part the left and right lead regions contain infinite repeated units; in the lower part the two CAP regions with finite repeated units can mimic the two semi-infinite leads. The complex potentials (imaginary part) in the CAP regions are indicated by two blue curves.

Since the CAP is energy independent, we may write the Green's function with CAP (Eq. (6a)) into the spectrum form (see the detailed derivation in Appendix A)

$$G^r(r,r') = \sum_k \frac{\psi_k(r)\Phi_k^*(r')}{E-\varepsilon_k}, \qquad (7)$$

where $\psi_k(r)$ and $\Phi_k(r)$ are the eigenfunctions of the following two non-hermitian Hamiltonians

$$(H_0 + \sum_{\alpha=L,R} W_\alpha)\psi_k = \varepsilon_k \psi_k, \qquad (8a)$$

$$(H_0 + \sum_{\alpha=L,R} W_\alpha^\dagger)\Phi_k = \varepsilon_k^* \Phi_k, \qquad (8b)$$

which satisfy the bi-orthonormal relation [26] and $H_0$ is the Hamiltonian without CAP.

Equation (7) can also be recast into the atomic basis ($\{\phi_i(r)\}$)

$$G_{m,n}^r = \sum_k \frac{\psi_{m,k}\Phi_{n,k}^*}{E-\varepsilon_k} \qquad (9)$$

where $G_{m,n}^r = <\phi_m(r)|G^r(r,r')|\phi_n(r')>$, $\psi_{m,k} = <\phi_m(r)|\psi_k(r)>$ and $\Phi_{n,k}^* = <\Phi_k^*(r')|\phi_n(r')>$. As the eigenvalue $\varepsilon_k$ is a complex number, it is natural to consider that Eq. (7) or Eq. (9) has some Lorentzian expansion form. To see this, we write the numerator and $\varepsilon_k$ into the real and imaginary parts: $\psi_{m,k}\Phi_{n,k}^* = A_{k,(m,n)}^R + iA_{k,(m,n)}^I$ and $\varepsilon_k = \Omega_k + iW_k$, then we have

$$G_{m,n}^r = \sum_k \frac{(A_{k,(m,n)}^R + iA_{k,(m,n)}^I)(E-\Omega_k + iW_k)}{(E-\Omega_k - iW_k)(E-\Omega_k + iW_k)}$$

$$= \sum_k \frac{[(A_{k,(m,n)}^R(E-\Omega_k) - A_{k,(m,n)}^I W_k] + i[A_{k,(m,n)}^R W_k + A_{k,(m,n)}^I(E-\Omega_k)]}{(E-\Omega_k)^2 + W_k^2}. \qquad (10)$$

We see that this Lorentzian form has a little difference from the standard one ($\frac{A_k}{(E-\Omega_k)^2 + W_k^2}$) as in Eq. (3) or in our previous papers [14,16]. However, we make some modifications to the residue calculations and TDDFT-NEGF can also be implemented. The details are given in Appendix B.

In practical calculations, we need the Lorentzian expansion for the self-energy matrix, which comes from the surface Green's function of a semi-infinite lead. Instead of calculating the system with two CAP regions and one device region (as shown in Figure 1), we may use one CAP region to mimic a semi-infinite lead. For example, the surface Green's function of the left lead is calculated as

$$g^r_{L,m,n} = \sum_k \frac{\psi_{L,m,k} \Phi^*_{L,n,k}}{E - \varepsilon_k} \tag{11}$$

where $\psi_{L,m,k}$ and $\Phi_{L,n,k}$ are the eigenvectors of the CAP-involved Hamiltonian: $\mathbf{H}_L + \mathbf{W}_L$ and its conjugate $\mathbf{H}_L + \mathbf{W}^\dagger_L$, respectively. ($\mathbf{H}_L$ and $\mathbf{W}_L$ are the same as those in Eq. (6); the subscript '$L$' denotes the left part of the lead.) We now focus on a larger system which contains the device and the left lead (CAP) regions. The self-energy matrix is evaluated as

$$\mathbf{\Sigma}^r_L = \mathbf{H}_{DL} \mathbf{g}^r_L \mathbf{H}_{LD} \tag{12}$$

where $\mathbf{H}_{DL}$ and $\mathbf{H}_{LD}$ are the coupling matrices between the device and the left CAP region. Substituting Eq. (11) into Eq. (12), the self-energy matrix is recast into the Lorentzian form

$$\Sigma^r_{L,m,n} = \sum_{m'n'} H_{mm'} g^r_{L,m',n'} H_{n',n} = \sum_k \frac{B_{L,k,(m,n)}}{E - \varepsilon_k} \tag{13}$$

where $B_{L,k,(m,n)} = \sum_{m'n'} H_{mm'} \psi_{L,m',k} \Phi^*_{L,n',k} H_{n',n}$.

Now we have shown that the self-energy matrix can be written into such Lorentzian form as well. So we find a natural Lorentzian expansion scheme for TDDFT-NEGF calculation.

## C. TDDFT-NEGF theory

TDDFT-NEGF theory solves the equations of motion (EOM) for the density matrix in an open system, based on the nonequilibrium Green's function theory. When the open system is partitioned into three regions of the left lead ($L$), device ($D$) and the right lead ($R$), the EOM for the device is given below

$$i\dot{\boldsymbol{\sigma}}_D(t) = [\mathbf{h}_D(t), \boldsymbol{\sigma}_D(t)] + \sum_{\alpha=1}^{N_\alpha} [\mathbf{h}_{D\alpha} \boldsymbol{\sigma}_{\alpha D}(t) - \boldsymbol{\sigma}_{D\alpha}(t) \mathbf{h}_{\alpha D}], \tag{14}$$

where $\boldsymbol{\sigma}_D(t)$ and $\mathbf{h}_D$ are the single-electron density matrix and Hamiltonian of the device. $\boldsymbol{\sigma}_{\alpha D}(t)$ ($\boldsymbol{\sigma}_{D\alpha}(t)$) and $\mathbf{h}_{\alpha D}$ ($\mathbf{h}_{D\alpha}$) are the coupling density matrix and the coupling Hamiltonian between the device $D$ and the lead $\alpha$ ($\alpha = L$ or $R$).

With the relation $\boldsymbol{\sigma}_D(t) = -i\mathbf{G}^<_D(t,t)$ and some derivations [9, 13], we have

$$i\dot{\boldsymbol{\sigma}}_D(t) = [\mathbf{h}_D(t), \boldsymbol{\sigma}_D(t)] + i \sum_{\alpha=1}^{N_\alpha} \int_{-\infty}^t d\tau [\mathbf{G}^<_D(t,\tau) \cdot \mathbf{\Sigma}^>_\alpha(\tau,t) - \mathbf{G}^>_D(t,\tau) \cdot \mathbf{\Sigma}^<_\alpha(\tau,t) + H.C.] \tag{15}$$

where $\mathbf{\Sigma}^x_\alpha(t,\tau)$ is the lesser ($x=<$) or greater ($x=>$) self-energy for the lead $\alpha$; $\mathbf{G}^x_D(t,\tau)$ is the lesser or greater Green's function of the device. $H.C.$ means the Hermitian conjugate. The current between the lead and the device is evaluated similarly [16]

$$J_\alpha(t) = -2eTr\{\text{Re}(\int_{-\infty}^{t} d\tau [\mathbf{G}_D^<(t,\tau) \cdot \mathbf{\Sigma}_\alpha^>(\tau,t) - \mathbf{G}_D^>(t,\tau) \cdot \mathbf{\Sigma}_\alpha^<(\tau,t)])\}. \tag{16}$$

where $Tr$ is the trace operator.

Equation (15) is difficult to be solved, since the lesser or greater Green's function is related to the retarded and advanced Green's functions, and $\mathbf{G}_D^r(t,t_1)$ has to be solved from the differential-integral equation [11]. Several algorithms for solving the EOM of $\mathbf{\sigma}_D(t)$ were proposed [11-12]. We opt for another method. Instead of solving the EOM of the Green's functions, new matrices are defined as follows:

$$\mathbf{\varphi}_\alpha(\varepsilon,t) = i\int_{-\infty}^{t} d\tau [\mathbf{G}_D^<(t,\tau) \cdot \mathbf{\Sigma}_\alpha^>(\varepsilon,\tau,t) - \mathbf{G}_D^>(t,\tau) \cdot \mathbf{\Sigma}_\alpha^<(\varepsilon,\tau,t)], \tag{17}$$

$$\mathbf{\varphi}_{\alpha\alpha'}(\varepsilon,\varepsilon',t) = i\int_{-\infty}^{t} dt_1 \int_{-\infty}^{t} dt_2 \{[\mathbf{\Sigma}_{\alpha'}^<(\varepsilon',t,t_1) \cdot \mathbf{G}_D^a(t_1,t_2) + \mathbf{\Sigma}_{\alpha'}^r(\varepsilon',t,t_1) \cdot \mathbf{G}_D^<(t_1,t_2)]\mathbf{\Sigma}_\alpha^>(\varepsilon,t_2,t)$$

$$-[\mathbf{\Sigma}_{\alpha'}^>(\varepsilon',t,t_1) \cdot \mathbf{G}_D^a(t_1,t_2) + \mathbf{\Sigma}_{\alpha'}^r(\varepsilon',t,t_1) \cdot \mathbf{G}_D^>(t_1,t_2)]\mathbf{\Sigma}_\alpha^<(\varepsilon,t_2,t)\}, \tag{18}$$

where $\mathbf{\Sigma}_\alpha^{<,>}(\varepsilon,\tau,t)$ is the energy resolved self-energy: $\mathbf{\Sigma}_\alpha^{<,>}(\tau,t) = \int d\varepsilon \cdot \mathbf{\Sigma}_\alpha^{<,>}(\varepsilon,\tau,t)$.

$\mathbf{\varphi}_\alpha(\varepsilon,t)$ and $\mathbf{\varphi}_{\alpha\alpha'}(\varepsilon,\varepsilon',t)$ are termed as the 1st and 2nd tier energy dispersed dissipation matrices, respectively. With the EOM of self-energies and the Green's functions, the time derivatives of $\mathbf{\sigma}_D(t)$, $\mathbf{\varphi}_\alpha(\varepsilon,t)$ and $\mathbf{\varphi}_{\alpha\alpha'}(\varepsilon,\varepsilon',t)$ can be derived, which are given in the references [13] and [14]. These differential equations constitute a closed set of hierarchical equations, which is exact and solvable.

In practical calculation for these equations, both the energy integration and the multiple energy components of dissipation matrices lead to huge computation. Some simplification has to be made by transforming the energy integration into the some summation. The details can be found in our previous papers [14]. The equations of motion for $\mathbf{\sigma}_D(t)$, $\mathbf{\varphi}_\alpha(\varepsilon,t)$ and $\mathbf{\varphi}_{\alpha\alpha'}(\varepsilon,\varepsilon',t)$ can be recast in the following discrete form:

$$i\dot{\mathbf{\sigma}}_D(t) = [\mathbf{h}_D(t),\mathbf{\sigma}_D(t)] - \sum_\alpha \sum_{k=1}^{N_k} (\mathbf{\varphi}_{\alpha k}(t) - \mathbf{\varphi}_{\alpha k}^\dagger(t)), \tag{19}$$

$$i\dot{\mathbf{\varphi}}_{\alpha k}(t) = [\mathbf{h}_D(t) - i\gamma_{\alpha k}^+ - \mathbf{\Delta}_\alpha(t)]\mathbf{\varphi}_{\alpha k}(t) - i[\mathbf{\sigma}_D(t)\mathbf{A}_{\alpha k}^{>+} + \bar{\mathbf{\sigma}}_D(t)\mathbf{A}_{\alpha k}^{<+}] + \sum_{\alpha'}\sum_{k'=1}^{N_k} \mathbf{\varphi}_{\alpha k,\alpha' k'}(t), \tag{20}$$

$$i\dot{\mathbf{\varphi}}_{\alpha k,\alpha' k'}(t) = -[i\gamma_{\alpha k}^+ + \mathbf{\Delta}_\alpha(t) + i\gamma_{\alpha' k'}^- - \mathbf{\Delta}_{\alpha'}(t)] \cdot \mathbf{\varphi}_{\alpha k,\alpha' k'}(t)$$

$$+ i(\mathbf{A}_{\alpha' k'}^{>-} - \mathbf{A}_{\alpha' k'}^{<-})\mathbf{\varphi}_{\alpha k}(t) - i\mathbf{\varphi}_{\alpha' k'}^\dagger(t)(\mathbf{A}_{\alpha k}^{>+} - \mathbf{A}_{\alpha k}^{<+}). \tag{21}$$

where $\bar{\mathbf{\sigma}}_D = \mathbf{1} - \mathbf{\sigma}_D$, $\mathbf{A}_{\alpha,k}^{<,>,\pm}$ and $\gamma_{\alpha,k}^\pm$ are from the residue calculations. For the CAP, these residue results are given in Appendix B, which are different from those in Ref. [14]. $\mathbf{\varphi}_{\alpha k}(t)$ and $\varphi_{\alpha k,\alpha'k'}(t)$ are the discrete versions of 1st tier and 2nd tier energy dispersed dissipation matrices. They are defined similarly as Eqs. (17) and (18), but the energy resolved self-energy $\Sigma_\alpha^{<,>}(\varepsilon,\tau,t)$ is replaced by the discrete $\Sigma_{\alpha,k}^{<,>}(\tau,t)$, which are obtained from the following integral-summation transformation

$$\Sigma_\alpha^{<,>}(\tau,t) = \int d\varepsilon \cdot \Sigma_\alpha^{<,>}(\varepsilon,\tau,t) = \sum_{k=1}^{N_k} \Sigma_{\alpha k}^{<,>}(\tau,t). \tag{22}$$

The numerical procedure of TDDFT-NEGF method is summarized as follows,
1. The Hamiltonian is constructed from the equilibrium Kohn-Sham Fock matrix of the self-consistent field calculation (the first principles model) or from the tight-binding model.
2. The self-energies of the leads are approximated by the multi-Lorentzian expansion from the CAP method or from the least square method.
3. The initial state of Eqs.(19)-(21) is calculated by the residue calculation method as stated in literatures [16,17].
4. The fourth-order Runge-Kutta scheme is used to solve TDDFT-NEGF equations (Eqs. (19-21)) and thus the transient current is obtained (Eq. (16)).

### III. Results and discussions

In this section, we test the CAP method for a simple system with the tight-binding model. Then we apply the CAP method to simulate the graphene nanoribbon systems and use the CAP-based Lorentzian expansion in the TDDFT-NEGF calculation. At last we turn to a first- principles model: the carbon-chains with a benzene molecule and carry out the time dependent quantum transport calculation with the CAP method.

A. 1D atom chain system (TB model)

The system is a 1D-atom chain which is modeled by the nearest neighbor tight-binding (TB) Hamiltonian. Each atom has one orbital. The hopping matrix element $t$ is 2.7 eV. There are 2 atoms in the device region. In the two lead regions, each has $N_L$ repeated sites with the CAP potential. Since we chose the TB model, only the diagonal terms of $W_{\alpha,mn}$ in Eq. (5) are calculated. The Green's function is obtained from Eq. (6a) and the local density of states (LDOS) of the device system is obtained by $\rho_i = \frac{-1}{\pi} \text{Im}[G_{i,i}^r(E)]$. For a homogeneous infinite 1D-atom chain, the LDOS may also be calculated analytically from the iteration solution of the Dyson's equation [2]:

$$\rho(E) = \frac{1}{\pi}(\frac{1}{2t})\frac{1}{\sqrt{1-(\frac{E-\varepsilon_0}{2t})^2}}. \tag{23}$$

Figure 2(a) shows that when the CAP range is long enough (larger than 10 repeated units), the LDOS curve from the CAP calculation (solid line) is very close to the accurate NEGF result from Eq. (6b) (dashed line). Figure 2(b) shows the transmission spectra. Similarly when the CAP region is long enough, the transmission spectrum from the CAP calculation (solid line) is very close to the accurate result (dashed line). The transmission is obtained from the following formula:

$$T = Tr[\mathbf{\Gamma}_L \mathbf{G}^r \mathbf{\Gamma}_R \mathbf{G}^a] \tag{24}$$

where $\mathbf{\Gamma}_L = 2\,\text{Im}[\mathbf{\Sigma}_L^r]$, $\mathbf{\Gamma}_R = 2\,\text{Im}[\mathbf{\Sigma}_R^r]$, $\mathbf{\Sigma}_L^r$ and $\mathbf{\Sigma}_R^r$ are the self-energies obtained iteratively in NEGF calculation. Eq. (24) can also be used for the CAP calculation, in which case the self-energy is calculated by Eq. (13).

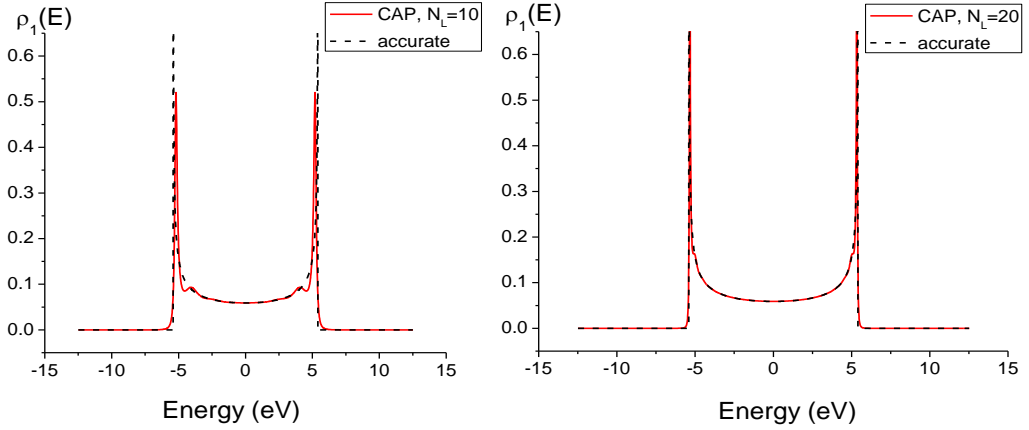

(a)

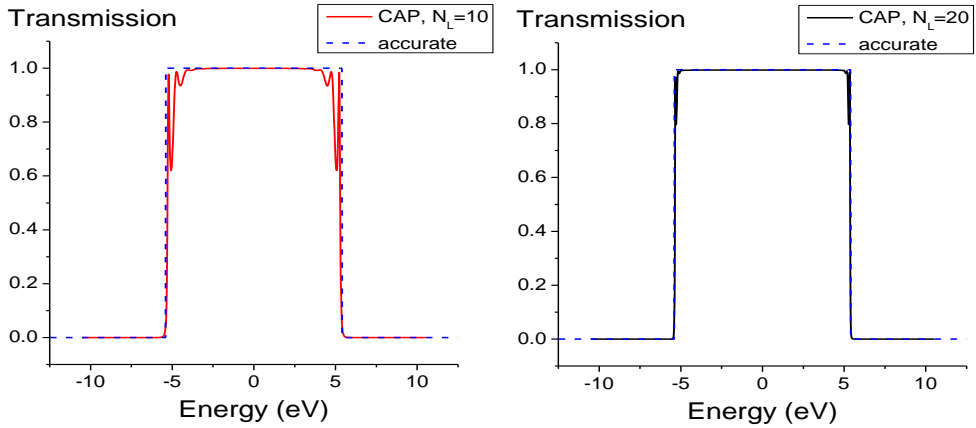

(b)

FIG. 2 (a) The LDOS curves in a 1D atom-chain system from the CAP (solid line) and the accurate (dashed line) calculations. (b) The transmission spectrum from the CAP (solid line) and the accurate (dashed line) calculations. In (a) and (b) the CAP regions include 10 repeated units in the left figures and 20 repeated units in the right figures.

B. Graphene nanoribbon system (TB model)

Now we examine the zigzag graphene nanoribbon (ZGNR) systems. We focus first on a uniform ZGNR with M=8 atoms in one unit (see Fig. 3 (a)). The nearest neighbor tight-binding model is used. The hopping integral in the TB model is set to -2.7 eV.

In general, we need two sets of the eigenvalues (for the left and right CAP regions) for the Lorentzian expansion. If the two leads are identical, it seems that only one set of Lorentzian is enough. However, in such case we find that with a common partition, both left leads and right leads often have different sets of CAP eigenvalues. This is because that although the artificial absorbing potentials ($\mathbf{W}_L$ and $\mathbf{W}_R$) are mirror symmetric for the two identical leads, the Hamiltonians of two leads ($\mathbf{H}_L$ and $\mathbf{H}_R$) may not have such a symmetry. This leads to the different eigenvalues. This can be seen more clearly in Fig. 3. Figure 3(a) shows the commonly partitioned lead-device-lead system with three repeated units in the left and right lead parts. The Hamiltonian of left (or right) CAP regions with 15 repeated units is constructed from the Hamiltonian of the lead part with 3 repeated units. The dimension of the left (or right) CAP Hamiltonian is 120. Since $\mathbf{H}_L$ and $\mathbf{H}_R$ have no mirror symmetry, $\mathbf{H}_L + \mathbf{W}_L$ and $\mathbf{H}_R + \mathbf{W}_R$ have no identical eigenvalues, as shown in Fig. 3(c).

Alternatively, we can partition and index the left and right leads symmetrically as shown in Fig. 3(b): the geometry and Hamiltonians of left and right CAP regions are of mirror symmetric. So the left and right CAP regions have the same eigenvalues (Fig. 3(d)) and only 120 Lorentzians are needed in the TDDFT-NEGF calculation.

Since in the TDDFT-NEGF calculation, a large number of Lorentzians (denoted by $N_d$) will lead to a large size of auxiliary density matrices and heavy computation lode, it is necessary to reduce $N_d$ value further. We may use a combination scheme to reduce the number of Lorentzian points in the $W$ v.s. $\Omega$ plot ($W$ and $\Omega$ are the width and center of the Lorentzian functions). This scheme combines the Lorentzians with the closed $W$ and $\Omega$ values into a single one. The details of this scheme are given in our previous paper [16]. One example for this zigzag graphene ribbon is shown in Fig. 3(e): the original 120 Lorentzian points (from the symmetric partition) are combined into 57 new points. With these combined Lorentzian points, we obtain the new amplitudes ($B_{L,k,(m,n)}$ in Eq. (13)) by fitting all the self-energy curves. By the NEGF calculation, the final transmission spectrum calculated with these combined Lorentzians also agrees well with the accurate one, as shown in Fig. 3(f).

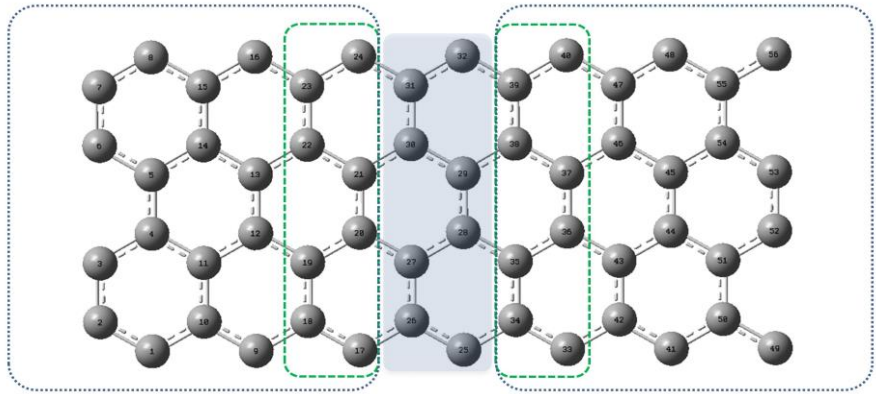

(a)

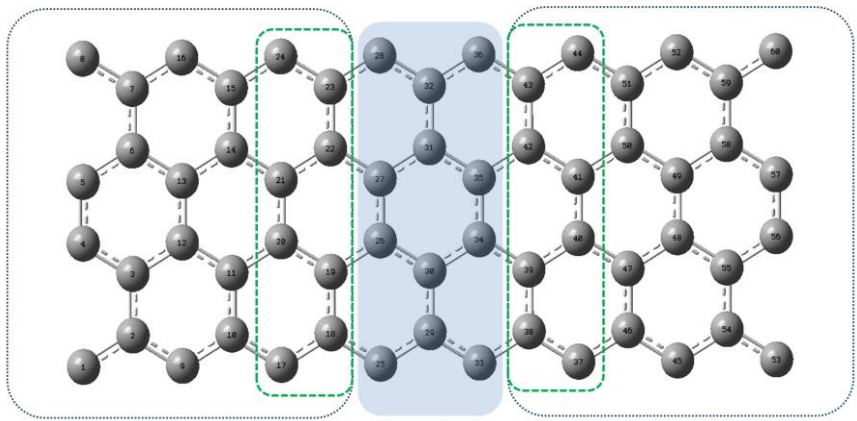

(b)

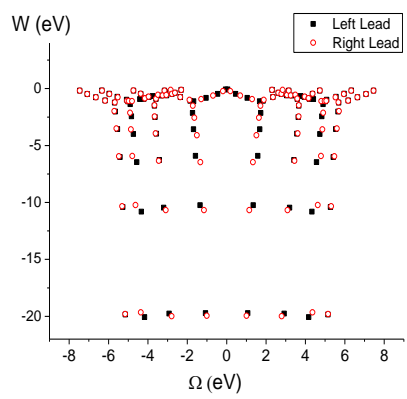

(c)

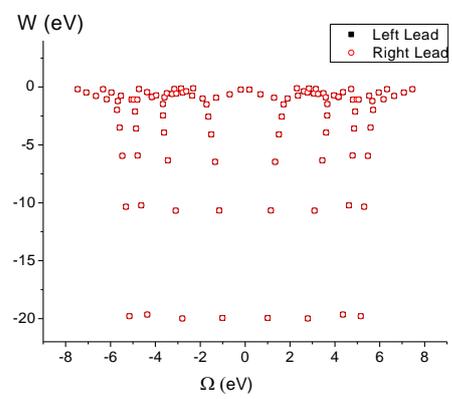

(d)

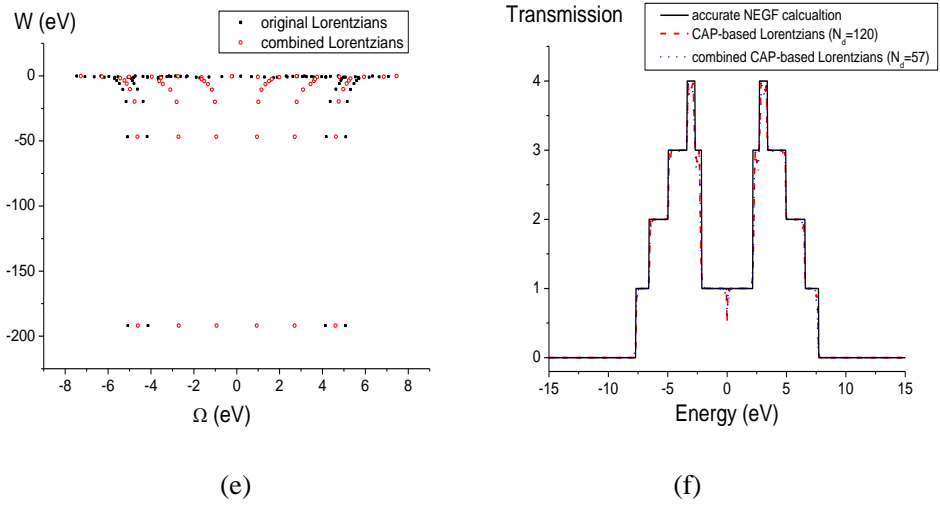

(e)                  (f)

FIG. 3 (a) and (b): the atomic structure and index for the ZGNR (M=8). The middle rectangle is for the device part and the left and right rectangles are for the lead parts. Each lead contains 3 repeated units (the first unit is indicated by a dashed rectangle). In (a) ZGNR is partitioned asymmetrically and in (b) it is partitioned mirror symmetrically for the left and right leads. (c) The real and imaginary part of eigenvalues (expressed as $\Omega$ and $W$ as shown in Eq. (10)) in the CAP regions with the asymmetric partition. The eigenvalues in the left (filled circle) and right (empty circle) CAP regions are not the same. (d) The real and imaginary part of eigenvalues from the CAP regions with the symmetric partition. The eigenvalues in the left (filled circle) and right (empty circle) CAP regions are identical. (e) The original Lorentzian points and the combined Lorentzian points in the $W$ v.s. $\Omega$ plot. The number of Lorentzians is reduced from 120 to 57 with a combination radius of 1.0 eV. It is noted that (c) and (d) only show part of the Lorentzian points for clarity. (f) The transmission spectrum of the zigzag graphene nanoribbons with M=8 atoms in one unit. The solid line is from the accurate NEGF calculation; the dashed line is from the CAP-based Lorentzian calculation (with 120 Lorentzians) and the dotted line is from the combined Lorentzian calculation (with 57 Lorentzians).

Then we use this symmetric partition strategy to study the dynamic transport of a combined GNR system. We choose the following structure as an example: the two leads are zigzag GNRs (M=8) and the device part is another smaller zigzag GNR (M=6). The Lorentzian expansion (with 160 Lorentzians from 20 repeated units) is obtained from the CAP calculation as stated previously. 50 Padé points is used in the Padé spectrum decomposition (see Eq. (2)) [24]. Figure 4(a) shows the atomic structure and parition scheme for this combined GNR system. Figure 4(b) shows the transmission spectrum of this system. We see the spectrum shape is similar to the pure ZGNR case (Fig. 3(f)). There are some oscillations in the middle part of the spectrum due to the interference effect between the device-lead interfaces. The steady state solution of TDDFT-NEGF is obtained from the rapid residue calculation method developed in our previous papers [16-17]. Then the TDDFT-NEGF simulation is implemented with the 4[th] order Runge-Kutta scheme for solving Eqs. (19)-(21) numerically. Figure 4(c) shows the dynamic currents through the left lead of this system. A bias volatge with exponentially change ($\Delta(t) = V_0(1-\exp[-t/\tau])$) is applied symmetrically on the device. The on-site energies of the device Hamiltonian changes linearly between two leads. From the figure we see that there exist large oscillations in the beginning, which is the over-shotting effect. This is due to the very narrow spectrum of the lead [16]: Only a small amount of electron near the Fermil level can be disspated into the GNR lead, so most of the electron wave injected in the device is reflected on the device-lead boundary, which gives the oscillation current. We calculate the currents for rapidly-rising bias (solid line, $\tau = 0.01$ fs) and slowly-rising bias (dashed line, $\tau = 0.5$ fs). The rapidly-rising bias causes much larger over-shooting current while another bias causes smaller overshooting current. This is resonable since for the slowly-rising bias, the injected electron has enough time to escape into the right lead.

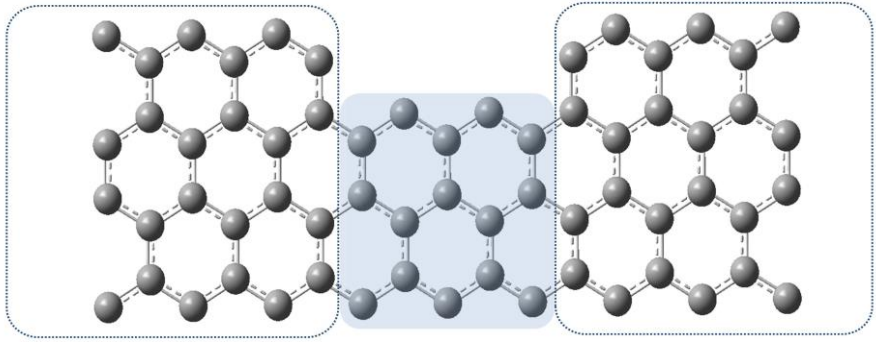

(a)

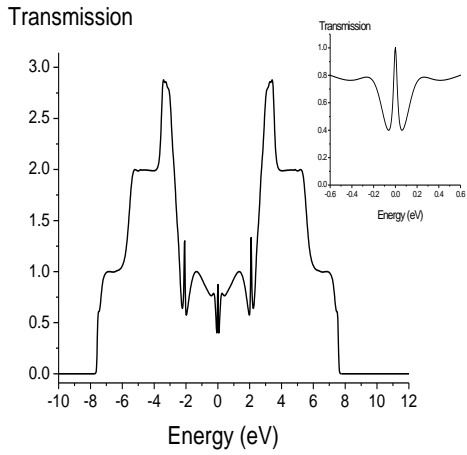

(b)

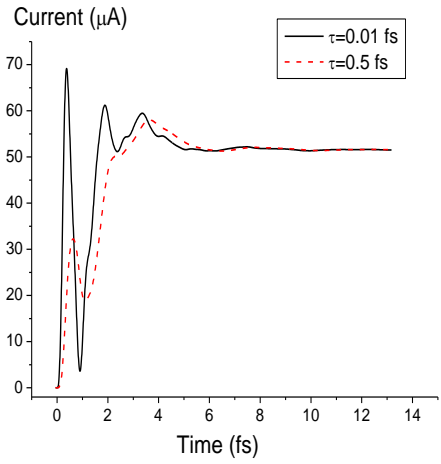

(c)

FIG. 4 (a) The atomic structure for the ZGNR (M=8)-ZGNR(M=6)-ZGNR(M=8) system. The middle shadow indicates the device region and the two rectangles indicate the two lead regions. (b) The transmission spectrum of this composite GNR system. The inset is the magnified part near the Fermi energy. (c) The dynamic current of this composite system under a bias voltage $\Delta(t) = V_0(1-\exp[-t/\tau])$ applied symmetrically on the two device sides, where $V_0 = 1$ V, the solid line is for the rapidly-rising case ($\tau = 0.01$ fs) and the dashed line is for the slowly-rising case ($\tau = 0.5$ fs).

C.  Carbon-chain-benzene-carbon-chain system (DFTB model)

Now we turn to the realistic system modeled by the density-functional-based-tight-binding (DFTB) Hamiltonian [27-28]. This model is an approximation of DFT method derived from the second-order expansion of DFT Kohn-Sham energy around the reference charge density. The minimal basis set STO-3G is used, and adiabatic local-density approximation (ALDA) is adopted as the XC functional. The in-house software '*LODESTAR*' is used to generate the Hamiltonian and overlap matrix [29-30].

To test the CAP method, we first calculate the carbon-atom-chain system. This system is a uniform carbon-atom chain arrayed in the *x* direction with a distance of 1.4 angtrom. This 1D carbon chain has been investigated both in theory and experiments [31-32]. Here we choose a cumulene-type chain as a simple example in our CAP calculations, in which all carbon atoms are connected by double bonds. In the DFTB model, each carbon atom has 4 orbitals (s, $p_x$, $p_y$ and $p_z$). As TDDFT-NEGF theory requires orthogonal basis, an orthogonalization procedure is employed to transform the non-orthogonal atomic orbitals into the orthogonal ones [33-34] (also see details in Appendix C). These orthogonalized bases remain the local property. The local property is very important in transport calculations because if the new basis is spatially distributed, the lead-device partition would become meaningless. In Fig. 5(a), the upper panel shows the transformation matrix for several orbitals by this symmetric transformation, which exhibits how the new bases are constructed from the original atomic ones. This figure indicates the new bases are also locally positioned. The lower panel of Fig. 5(a) shows the transformation matrix by another canonical orthogonalization method [33], which indicates that the new bases extend in the whole space. So this type of orthogonalization is not suitable for partition in the transport calculation.

Since there exist nonzero Hamiltonian matrix elements between the orbitals of different neighbor atoms, we set 10 carbon atoms as one unit in the device and CAP regions. We choose 3 repeated units in the left (or right) CAP regions and 1 repeated unit in the device region. In the CAP calculation, $\mathbf{W}_\alpha$ changes on each atom instead of each unit. This gradual change of CAP makes the number of repeated units greatly decreased from 10-20 to 2-3 and the transmission spectrum still remains as good as that from the accurate result (see Fig. 5(b)).

However, when we use the eigenvector expansion scheme (Eq. (11) and (13)) to calculate the self-energy and the transmission spectrum, we find that in the energy range from -10 eV to 2 eV, the transmission curve deviates greatly from the accurate one. To find the reason, we draw the LDOS curves of the 4 orbitals, as shown in Fig. 5(c). We see that $p_y$ and $p_z$ orbitals contribute to the LDOS and the transmissions in the energy range from -10 eV to 2 eV. It indicates the problems lies in these $p_y$ and $p_z$ orbitals. We further notice that there exit degenerate eigenvalues (in Eq. (11)) for the CAP region which come from the degeneracy of $p_y$ and $p_z$ orbitals (they are equivalent due to the geometry of this 1D-chain). Finally, we find that for the two degenerate eigenvalues, their corresponding eigenvectors are not orthogonal to each other, which causes the fails of eigenvector expansion for the Green's function (Eq. (9)).

To fix this problem, we may orthogonalize all the eigenvectors to obtain the right Green's function. Alternately, another simple way can be utilized: we modify the CAP in y and z directions to eliminate the $p_y$-$p_z$ degeneracy. For this DFTB model, we make the diagonal element of each $p_y$ orbitals have some difference from that of the $p_z$ orbital. Using this new anisotropic CAP, the Green's function is calculated rightly and the transmission spectrum agrees very well to the accurate one, as in Fig. 5(b).

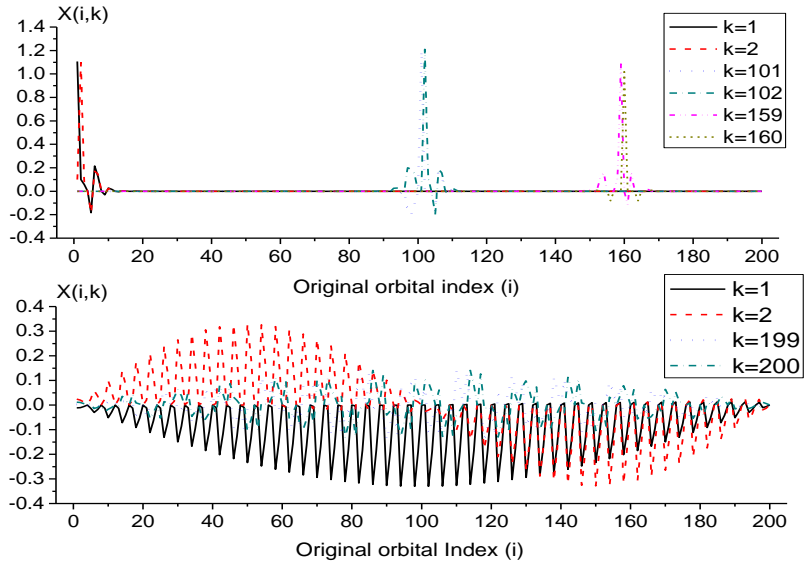

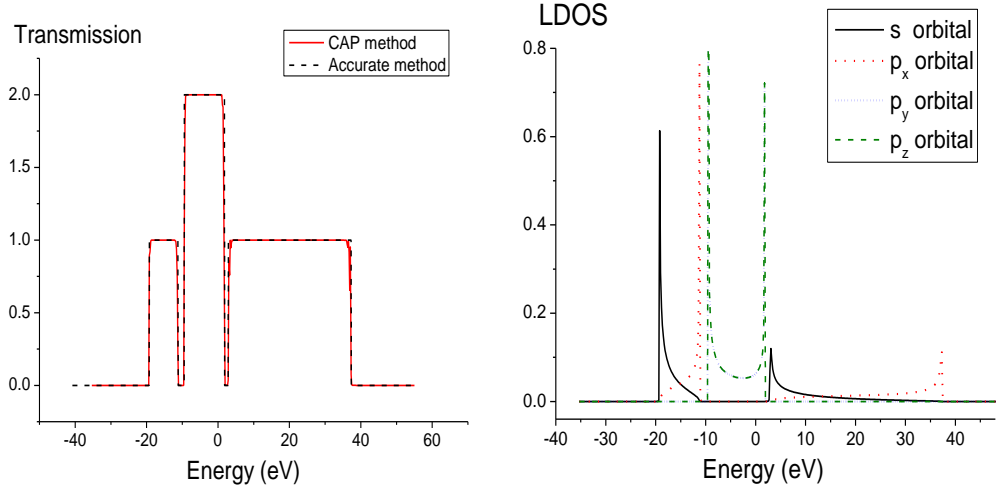

FIG. 5 (a) The local property of the orthogonalization transformation for a carbon-atom-chain system (with 50 atoms or 200 orbitals in DFTB model). The upper panel is from the symmetric orthogonalization and the lower panel is from the canonical orthogonalization method. (b) The transmission spectrum of a carbon-atom-chain system obtained from the CAP (solid line) and the accurate calculation (dashed line). 2 repeated units are used for the CAP calculation and the CAP varies gradually in each atom. (c) The local density of states curves for s, $p_x$, $p_y$ and $p_z$ orbitals on each atom in an open carbon-atom-chain system.

Now we come to a system with a benzene molecule sandwiched by two carbon chains, as shown in Fig. 6(a). The Hamiltonian is extracted from a much larger isolate system (with a device part and two long lead parts) calculated by the software '*LODESTAR*'. Then it is transformed into the orthogonal basis set as stated before. The Fermi level of this system lies at -5.088 eV. As stated before, three repeated units of carbon atoms are used for the CAP calculation. There are 120 orbitals in the CAP region and 120 Lorentzians are generated for the following calculation. Figure 6(b) shows the transmission spectra of this system calculated from the CAP method (solid line) and the accurate NEGF method (dashed line). They agree well with each other. From the lead spectrum of the carbon chain (see Fig. 3(b) in Reference [17]) we see that it is very flat near the Fermi level, thus the carbon-atom-chain behaves like a wide-band-limit (WBL) lead near the Fermi energy [17]. Compared to the sharp spectrum of the zigzag GNR lead (see Fig. 4(c) in Reference [16]), this flat spectrum also results in a much weaker over-shooting behavior, as shown in Fig. 6(c).

With these 120 Lorentzian expansion terms and 50 Padé decomposition terms, the steady and dynamic TDDFT-NEGF calculations are employed. In Fig. 6(c) the dynamic currents are induced by a bias voltage symmetrically applied on the two leads. We see in the long-time limit all the dynamic currents approach to the steady-state values (the horizontal dashed lines) calculated by the Landauer formula. For $V_0$=1.0V, the dynamic current with small $\tau$ (0.01 fs) (solid line) exhibits a lot of high-frequency oscillations than that with large $\tau$ (0.1 fs) (dashed line). This can be explained as follows: the rapidly-rising bias voltage (corresponding to small $\tau$) has a very wide spectrum from the Fourier transformation, which contains a lot of high-frequency components. So this rapid bias can induce a lot of high frequency currents, as the small oscillations in the current curve.

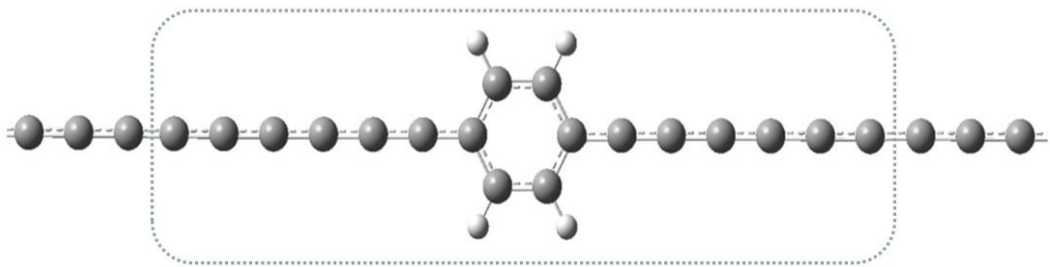

(a)

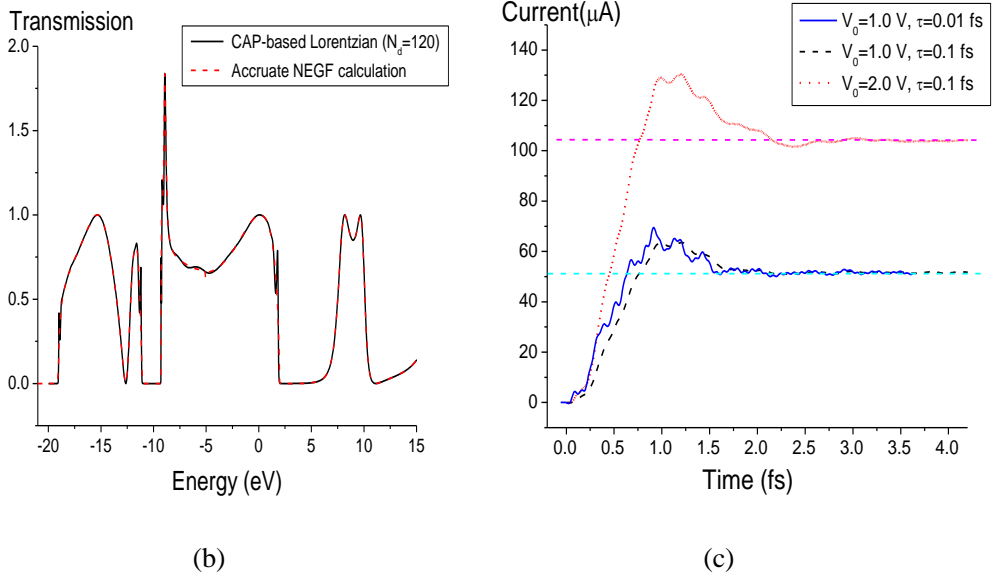

(b)                                       (c)

FIG. 6 (a) The atomic structure for the carbon-chain-benzene-carbon-chain system. The rectangle box indicates the device region. (b) The transmission spectrum of this carbon chain with benzene system. The solid line is from the CAP calculation and the dashed line is from accurate NEGF calculation. (c) The dynamic current through the right lead of this C-Benzene-C system. A bias voltage $\Delta(t) = V_0(1-\exp[-t/\tau])$ is symmetrically applied on the two leads and the potential in the device changes linearly between the leads. The three curves are for different parameters (the dotted line: $V_0 = 2.0V$, $\tau = 0.1$ fs; the dashed line: $V_0 = 1.0V$, $\tau = 0.1$ fs and the solid line: $V_0 = 1.0V$, $\tau = 0.01$ fs). The two horizontal lines are for the steady-state currents.

## IV. Conclusion

The Lorentzian expansion form of the surface Green's function and the self-energy matrix is derived with the CAP method. This method, based on mimicking the infinite environment with a finite absorbing region, generates the self-energy in Lorentzian forms for any open electronic system. With this CAP-based Lorentzian expansion, the modified residues of the lesser and greater self-energies are derived for the time-dependent quantum transport calculation.

In the GNR calculations, a mirror symmetric partition scheme is proposed to reduce the number of Lorentzian terms for the system with identical leads. The over-shooting current of a composite GNR system is investigated by the TDDFT-NEGF calculation. In the carbon-atom-chain system, the degenerated eigenvectors are eliminated by the anisotropic CAP scheme. And the transient current of a carbon-chain-benzene-carbon-chain system is obtained by the TDDFT-NEGF calculation. The current response with different rising-times of the bias voltage is analyzed.

This CAP-based expansion is an efficient and accurate way to decompose the leads' self-energies in the TDDFT-NEGF calculation. However, for the large systems such as the silicon nanowire and large carbon nanotubes, the number of Lorentzians is still quite large for first-principle calculations. We find that the LS method may generate fewer Lorentzians for these systems, but many fitting parameters have to be adjusted case by case. We also find the number of Lorentzians resulted from the CAP method can be further reduced by combining several Lorentzians with similar energies, which is widely utilized in the LS method. Further work to effectively generate the Lorentzian terms will be pursued.

**Acknowledgements**

Support from the Hong Kong Research Grant Council (HKU 700912P, HKU 700711P, HKU700909P, HKUST9/CRF/11G) and AoE (AOE/P-04/08) are gratefully acknowledged.

**Appendix**

A. Spectrum expansion for the Green's function of a non-Hermitian system

It is easy to see that the Green's function can be written as the following spectrum form

$$G^r(r,r') = \sum_k \frac{\psi_k(r)\psi_k^*(r')}{E - \varepsilon_k}$$

where $\psi_k(r)$ is the eigenfunctions of the Hamiltonian $H_0$: $H_0\psi_k = \varepsilon_k\psi_k$. Different eigenfunctions obey the orthonormal relation: $\int \psi_k(r)\psi_l^*(r)dr = \delta_{k,l}$. But in open systems, we have to include the non-Hermitian self-energy $\Sigma_\alpha^r$ (or CAP term) into the Hamiltonian and the total Hamiltonian is not Hermitian. Thus the spectrum form of Green's function above is not valid in this case.

Here we introduce the concept of bi-orthonormal bases. We chose the discrete basis for the following discussion, where Hamiltonian and eigenfunctions are changed to matrix and eigenvectors. If there exist two types of eigenvectors ($\mathbf{e}_m$ and $\mathbf{f}_m$) for a non-hermitian matrix $\mathbf{H}_\Sigma$:

$$\mathbf{H}_\Sigma \cdot \mathbf{e}_m = \lambda_m \mathbf{e}_m \tag{A1}$$

$$\mathbf{H}^\dagger_\Sigma \cdot \mathbf{f}_n = \mu_n \mathbf{f}_n \tag{A2}$$

where † means transpose (*t*) conjugate (*) operation, it can be proved that $\mu_m = \lambda_m^*$, we can derive the following relation

$$\mathbf{f}_n^\dagger \cdot \mathbf{H}_\Sigma \cdot \mathbf{e}_m = \lambda_m \mathbf{f}_n^\dagger \cdot \mathbf{e}_m = [\mathbf{H}^\dagger_\Sigma \cdot \mathbf{f}_n]^\dagger \cdot \mathbf{e}_m = \lambda_n \mathbf{f}_n^\dagger \cdot \mathbf{e}_m$$

$$(\lambda_m - \lambda_n)\mathbf{f}_n^\dagger \cdot \mathbf{e}_m = 0.$$

From the equation above, it is easy to see that if $\lambda_m \neq \lambda_n$, $\mathbf{f}_n^\dagger \cdot \mathbf{e}_m = 0$ and if $\mathbf{e}_m$ is properly scaled, $\mathbf{e}_n^\dagger \cdot \mathbf{e}_n = 1$. Thus $\mathbf{e}_m$ and $\mathbf{f}_m$ obey the bi-orthonormal relation: $\mathbf{f}_n^\dagger \cdot \mathbf{e}_m = \delta_{n,m}$.

In the continue case, we see that the two sets of eigenfunctions $\psi_k(r)$ and $\Phi_k(r)$ in Eqs. (8a) and (8b) ($\mathbf{H}_\Sigma = \mathbf{H}_0 + \sum_\alpha \mathbf{W}_\alpha$) also obey the following bi-orthonormal relation:

$$\int \Phi_m(r)\psi_n^*(r)dr = \delta_{m,n}. \tag{A3}$$

These eigenfunctions constitute the complete basis. So we may find an expansion for the delta function:

$$\sum_n C_n \psi_n(r) = \delta(r-r').$$

With the bi-orthonormal relation above, it is easy to obtain: $C_n = \Phi_n^*(r')$, thus

$$\sum_n \psi_n(r)\Phi_n^*(r') = \delta(r-r'). \tag{A4}$$

For the retarded Green's function, the expansion form is written as

$$G^r(r,r') = \sum_k D_k \psi_k(r).$$

Substituting Eq. (A4) into the definition of $G^r(r,r')$, we have

$$(E-H_\Sigma)\sum_k D_k \psi_k(r) = \sum_n \psi_n(r)\Phi_n^*(r').$$

Considering Eq. (8a), the expansion coefficient is solved: $D_k = \dfrac{\Phi_k^*(r')}{E-\varepsilon_k}$. Thus the spectrum expansion form of Eq. (7) is obtained.

## B. Modified residues for self-energy matrices in TDDFT-NEGF

As we mentioned in Sec. IIC, since the Lorentzian function obtained from the CAP method is different from the standard Lorentzian form, the residue calculations in TDDFT-NEGF scheme have to be modified. The following shows the details.

The self-energy in steady state is given below

$$\overline{\boldsymbol{\Sigma}}_\alpha^x(\tau-t) = \frac{s_x i}{2\pi}\int_{-\infty}^{+\infty} f_\alpha^x(\varepsilon)\boldsymbol{\Lambda}_\alpha(\varepsilon)\cdot e^{-i\varepsilon(\tau-t)}d\varepsilon = \frac{s_x i}{2\pi}\oint f_\alpha(z)\boldsymbol{\Lambda}_\alpha(z)\cdot e^{-iz(\tau-t)}dz \tag{B1}$$

where $x = <,>$, $s_< = 1$, $s_> = -1$, $f_\alpha^<(z) = f_{P\alpha}(z)$, $f_\alpha^>(z) = 1 - f_{P\alpha}(z)$; $f_{P\alpha}$ is the Fermi function with the Padé decomposition: $f_{P\alpha}(z) = \dfrac{1}{2} + \sum_{p=1}^{N_p}(\dfrac{R_p/\beta}{z-\mu_\alpha-z_p^+/\beta} + \dfrac{R_p/\beta}{z-\mu_\alpha-z_p^-/\beta})$,

and $\beta = \dfrac{1}{k_B T}$; $\boldsymbol{\Lambda}_\alpha(z)$ is the linewidth function, which is related to the imaginary part of the retarded self-energy: $\boldsymbol{\Lambda}_\alpha(z) = -2\,\mathrm{Im}[\boldsymbol{\Sigma}_\alpha^r(z)]$. In the CAP-based Lorentzian expansion, $\boldsymbol{\Sigma}_\alpha^r(z)$ is expanded into the modified Lorentzian terms: $\boldsymbol{\Sigma}_\alpha^r(E) = \sum_k \dfrac{\mathbf{B}_{\alpha,k}}{E-\varepsilon_k}$ (Eq. (13)), we may write out the linewidth function as

$$\boldsymbol{\Lambda}_\alpha(E) = (-2)\cdot \sum_k^{N_d} \frac{\mathbf{B}_{\alpha,k}^R W_k + \mathbf{B}_{\alpha,k}^I (E-\Omega_k)}{(E-\Omega_k)^2 + W_k^2} \tag{B2}$$

where $\mathbf{B}_{\alpha,k}^{R}$ and $\mathbf{B}_{\alpha,k}^{I}$ is the real and imaginary part of $\mathbf{B}_{\alpha,k}$; $\Omega_k$ and $W_k$ is the real and imaginary part of the eigenvalue $\varepsilon_k$. In practical calculation, we have to ensure $W_k$ to be positive (due to the residues in different contours as stated below). For the retarded Green's function, all the singularities lie in the lower half complex plain, which means all $W_k$ are negative. So we make the transformation: $W_k' = -W_k$, and the linewidth function is given as

$$\Lambda_\alpha(E) = 2 \cdot \sum_{k}^{N_d} \frac{\mathbf{B}_{\alpha,k}^{R} W_k' - \mathbf{B}_{\alpha,k}^{I}(E-\Omega_k)}{(E-\Omega_k)^2 + W_k'^{2}} \ . \tag{B3}$$

With the residue theory, the integral in Eq. (B1) is transformed into the residue summation. It is noted that to ensure the integrant (or the factor $e^{-iz(\tau-t)}$) does not diverge on the integral contours, different contours have to be used depending on the sign of $\tau - t$. The details may be found in the literature [14]. The final residue results have the following form

$$\overline{\Sigma}_\alpha^{<,>}(\tau-t) = \sum_{k}^{N_k} \mathbf{A}_{\alpha k}^{<,>\pm} \cdot e^{\mp \gamma_{\alpha k}^{\pm}(t-\tau)} \ , \tag{B4}$$

where '+' and '−' in the superscripts correspond to different contours, due to the sign of $\tau - t$. The expressions for $\mathbf{A}_{\alpha k}^{<,>\pm}$ and $\gamma_{\alpha k}^{\pm}$ with the modified Lorentzians are calculated here.

(1) In the case of $\tau - t < 0$:

$$\mathbf{A}_{\alpha k}^{<,+} = \begin{cases} i[\mathbf{B}_{\alpha,k}^{R} - i\mathbf{B}_{\alpha,k}^{I}] f_{P\alpha}(\Omega_k + iW_k') & (1 \le k \le N_d) \\ -\frac{R_p}{\beta} \cdot \Lambda_\alpha(z_p^+/\beta + \mu_\alpha) & (N_d + 1 \le k \le N_k) \end{cases} , \tag{B5}$$

$$\mathbf{A}_{\alpha k}^{>,+} = \begin{cases} -i[\mathbf{B}_{\alpha,k}^{R} - i\mathbf{B}_{\alpha,k}^{I}] \cdot [1 - f_{P\alpha}(\Omega_k + iW_k')] & (1 \le k \le N_d) \\ -\frac{R_p}{\beta} \cdot \Lambda_\alpha(z_p^+/\beta + \mu_\alpha) & (N_d + 1 \le k \le N_k) \end{cases} , \tag{B6}$$

$$\gamma_{\alpha k}^{+} = \begin{cases} W_k' - i\Omega_k & (1 \le k \le N_d) \\ -i(z_p^+/\beta + \mu_\alpha) & (N_d + 1 \le k \le N_k) \end{cases} \tag{B7}$$

where $p = N_k - k$, $z_p^+$ is the singularity of Padé decomposition in the upper complex plane; $\Lambda_\alpha$ is defined in Eq. (B3).

(2) In the case of $\tau - t > 0$:

$$\mathbf{A}_{\alpha k}^{<,-} = \begin{cases} i[\mathbf{B}_{\alpha,k}^{R} + i\mathbf{B}_{\alpha,k}^{I}] f_{P\alpha}(\Omega_k - iW_k') & (1 \le k \le N_d) \\ \frac{R_p}{\beta} \cdot \Lambda_\alpha(z_p^-/\beta + \mu_\alpha) & (N_d + 1 \le k \le N_k) \end{cases} , \tag{B8}$$

$$\mathbf{A}_{\alpha k}^{>,-} = \begin{cases} -i[\mathbf{B}_{\alpha,k}^{R} + i\mathbf{B}_{\alpha,k}^{I}] \cdot [1 - f_{P\alpha}(\Omega_k - iW_k')] & (1 \leq k \leq N_d) \\ \dfrac{R_p}{\beta} \cdot \mathbf{\Lambda}_\alpha(z_p^-/\beta + \mu_\alpha) & (N_d + 1 \leq k \leq N_k) \end{cases}, \quad (B9)$$

$$\gamma_{\alpha k}^{-} = \begin{cases} W_k' + i\Omega_k & (1 \leq k \leq N_d) \\ i(z_p^-/\beta + \mu_\alpha) & (N_d + 1 \leq k \leq N_k) \end{cases}. \quad (B10)$$

where $z_p^-$ is the singularity of the Padé decomposition in the lower complex plane.

### C. Basis orthogonalization

In our original TDDFT-NEGF theory, the bases or the orbitals are orthogonal while in the DFTB or DFT calculations, the bases are non-orthogonal. One way to solve this problem is to modify the TDDFT-NEGF theory for the non-orthogonal bases, which is shown in our recent paper [17]. Another way is to orthogonalize the original basis with some basis transformation. Here we show the details of this transformation.

Firstly we diagonalize the overlap matrix S: $\mathbf{S} = \int dr \boldsymbol{\varphi}^t \boldsymbol{\varphi} = \mathbf{U}\mathbf{S}_\Lambda \mathbf{U}^\dagger$, where $\boldsymbol{\varphi}$ is the original non-orthogonal basis, $\mathbf{S}_\Lambda$ is the diagonal matrix. Then we may construct the following transformation matrix ($\mathbf{X}$):

$$\mathbf{X} = \mathbf{S}^{-1/2} = \mathbf{U}\mathbf{S}_\Lambda^{-1/2}\mathbf{U}^\dagger.$$

It is easy to see that with this transformation ($\boldsymbol{\varphi}' = \boldsymbol{\varphi}\mathbf{X}$), the new bases ($\boldsymbol{\varphi}'$) are orthogonal: $\mathbf{S}' = \int dr \boldsymbol{\varphi}'^t \boldsymbol{\varphi}' = \mathbf{X}^\dagger \mathbf{S} \mathbf{X}^? = \mathbf{I}$. This type of basis transformation is often called the symmetric orthogonalization or the Löwdin's orthogonalization [33-34].

It is noted that there also exist other similar transformation matrices to orthogonalize the original basis set, such as $\mathbf{X} = \mathbf{U}\mathbf{S}_\Lambda^{-1/2}$ and $\mathbf{X} = \mathbf{U}\mathbf{S}_\Lambda^{-1/2}\mathbf{U}$. Our calculations similar to Fig.3(a) indicate that only the symmetric transformations (such as $\mathbf{X} = \mathbf{U}\mathbf{S}_\Lambda^{-1/2}\mathbf{U}^\dagger$ and $\mathbf{X} = \mathbf{U}S_\Lambda^{-1/2}\mathbf{U}$) remains the local property as the original atomic basis.